\documentclass[journal]{IEEEtran}
\IEEEoverridecommandlockouts
\usepackage{cite}
\usepackage{makecell}
\usepackage[utf8x]{inputenc}
\usepackage{graphicx}
\usepackage{subfigure}
\usepackage{array}
\usepackage{booktabs}
\usepackage{balance}
\usepackage{url}
\usepackage[misc]{ifsym}
\usepackage[dvips]{color}
\usepackage{mathrsfs}
\usepackage[linesnumbered,ruled,vlined]{algorithm2e}
\usepackage{amsmath}
\usepackage{nccmath}
\usepackage{amsfonts,amssymb,bm}
\usepackage{amsthm}
\usepackage[dvipsnames]{xcolor}
\usepackage[noend]{algpseudocode}
\usepackage{overpic}
\usepackage{multirow}

\SetCommentSty{mycommfont}
\SetKwInput{KwInput}{Input}                
\SetKwInput{KwOutput}{Output}              

\usepackage[normalem]{ulem}
\useunder{\uline}{\ul}{}

\makeatletter
\renewcommand{\@pnumwidth}{1.75em}
\renewcommand{\@tocrmarg}{2.75em}
\makeatother

\newcolumntype{C}[1]{>{\centering\let\newline\\\arraybackslash\hspace{0pt}}m{#1}}

\begin{document}  
\title{Multi-Static UWB Radar-based Passive Human Tracking Using COTS Devices}
\author{Chenglong~Li,
		Emmeric~Tanghe,~\IEEEmembership{Member,~IEEE,}
		Jaron~Fontaine,
		Luc~Martens,~\IEEEmembership{Member,~IEEE,}\\
		Jac~Romme,
		Gaurav~Singh,
		Eli~De Poorter,
		and Wout~Joseph,~\IEEEmembership{Senior~Member,~IEEE}
\thanks{C. Li, E. Tanghe, L. Martens, and W. Joseph are with the WAVES group, Department of Information Technology, Ghent University-imec, 9052 Ghent, Belgium (e-mail: chenglong.li@ugent.be).}
\thanks{J. Romme and G. Singh are with imec-Netherlands, 5656 AE Eindhoven, The Netherlands.}
\thanks{J. Fontaine and E. De Poorter are with the IDLab group, Department of Information Technology, Ghent University-imec, 9052 Ghent, Belgium.}
}


\maketitle
\begin{abstract}
Due to its high delay resolution, the ultra-wideband (UWB) technique has been widely adopted for fine-grained indoor localization. Instead of active positioning, UWB radar-based passive human tracking is explored using commercial off-the-shelf (COTS) devices. To extract the time-of-flight (ToF) reflected by the moving person, the accumulated channel impulse responses (CIR) and the corresponding variances are used to train the convolutional neural networks (CNN) model. Particle filter algorithm is adopted to track the moving person based on the extracted ToFs of all pairs of links. Experimental results show that the proposed CIR- and variance-based CNN models achieve less than 30-cm root-mean-square errors (RMSEs). Especially, the variance-based CNN model is robust to the scenario changing and promising for practical applications.

\end{abstract}

\begin{IEEEkeywords}
Indoor localization, Internet-of-Things, passive tracking, ultra-wideband, channel impulse response.
\end{IEEEkeywords}

\IEEEpeerreviewmaketitle

\section{Introduction}
\IEEEPARstart{L}{ocation} awareness is an essential feature of the extensive applications in Internet-of-things (IoT). Especially, radio frequency (RF)-based passive sensing has attracted increasing attention recently due to avoiding devices attached to the users. This is appreciated in specific use cases, such as customer behavior analysis and intruder detection. Moreover, compared with the conventional vision-based methods, RF-based passive sensing is little affected by the poor visibility and has no privacy issue, which is desirable for practical use. Passive sensing is reminiscent of the radar systems firstly invented in the early 20th century. But as hardware advancement, it is also possible for the lower-power commercial devices, for example wireless fidelity (WiFi) \cite{Li2016,Xie2019,Gao2020}, millimeter-wave radios \cite{wei2015,Wu2020}, and ultra-wideband (UWB) \cite{Chang2009,Ledergerber2020}, to enable the abundant applications in the field of smart IoT.
\par
UWB has been widely adopted in RF devices for the past five years, e.g., smartphones and car keys, which spawns extensive research on UWB-based applications. \cite{Salmi2011} proposed the UWB-based human breathing motion tracking. The prototype was implemented in an anechoic chamber and achieved centimeter-level accuracy. In \cite{Li2012}, UWB radar was introduced for the through-the-wall human detection, which could be used for life signal identification after a catastrophe. Moreover, UWB passive sensing has also been applied for crowd counting \cite{Choi2021} and remote health monitoring \cite{Khan2020}, which are significant, especially in the case of social distancing during the pandemic. However, most UWB passive sensing applications are based on the vector network analyzer (VNA) or other dedicated setups with perfect system settings (e.g., very accurate synchronization, high sampling rate, etc.). Instead of the dedicated devices, \cite{Ledergerber2020,Moschevikin2016} adopted the low-cost and commercial off-the-shelf (COTS) UWB modules for human monitoring and tracking and achieved promising results.
\par 
This paper focuses on fine-grained passive human tracking based on COTS UWB devices (i.e., Decawave DW1000 \cite{DWM1000}). Two reflected ToF estimation methods based on the CIR- and variance-based convolutional neural networks (CNN) models have been proposed. The main contributions of this paper are as follows: (i) Instead of simply background subtraction, the proposed CNN models learn the difference between the background and dynamic scenarios intrinsically, which can effectively mitigate the interference from the background. (ii) According to experimental validation, the proposed methods can achieve mean accuracy less than 30 cm that outperforms the state-of-the-art method in \cite{Ledergerber2020}. (iii) We have investigated the generation ability of the proposed CNN models.

\section{UWB-based Passive Human Tracking}
\subsection{UWB Experiment and CIR Preprocessing}
\begin{figure}[t]
\centering
\setlength{\abovecaptionskip}{-0.15cm}
\setlength{\belowcaptionskip}{-0.1cm}
\includegraphics[width=0.479\textwidth]{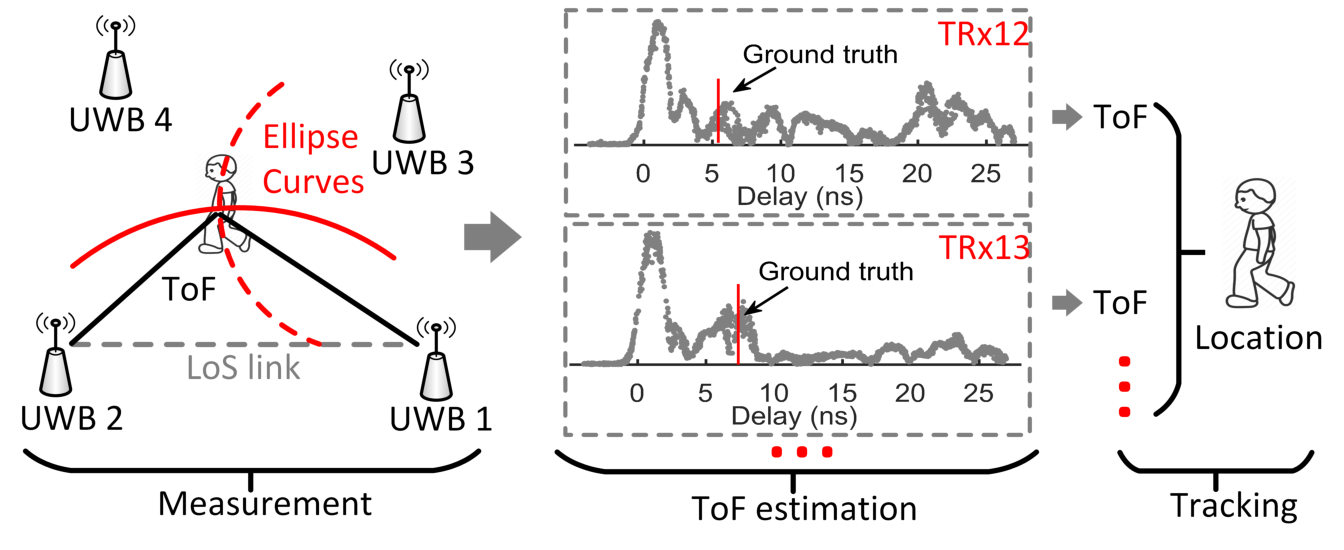}
\caption{Multistatic UWB radar-based passive human tracking.}
\label{fig:MethodStruc}
\end{figure}

The multi-static UWB radar-based human passive tracking experiment will be briefly introduced in this section. The corresponding dataset for passive human tracking is open-access and the detailed description can be found in \cite{Ledergerber2020,uwbDataset2020}. For the passive human tracking, the ToF reflected by the human body can be estimated based on the collected CIRs first. Then the locations of the moving person can be determined via the intersection of the multiple ellipse curves when we have the reflected ToFs of multiple UWB transceivers, as shown in Fig. \ref{fig:MethodStruc}. The UWB passive tracking experiment was conducted in an indoor laboratory environment with the moving range about 8$\rm{m}\times$6$\rm{m}$. Four UWB nodes with DWM1000 modules were deployed for the passive tracking. The ground truth of the moving trajectory was obtained via the motion capture (MoCap) system with millimeter-level accuracy. The pulse repetition frequency of DWM1000 is 16 MHz, the carrier frequency 3993.6 MHz, and the bandwidth 900 MHz. The chipping frequency is 499.2 MHz, so the time resolution of each CIR sample is 1.0016 ns. The total measurement duration was about two minutes, of which the first 17s and the last 16s indicate the case of no person moving. But at the beginning and end of the dynamic measurement, the target's ground truth measured by the MoCap system is not stable. So we denote the period between 21.4 s and 100 s as the dynamic scenario.
\par
In the dataset, for each single CIR measurement, only 31 samples have been reported, of which the fourth sample is the first peak (line-of-sight link) identified by DWM1000 using the leading edge (LDE) algorithm \cite{DWM1000}. As mentioned above, the sampling interval of DWM1000 is about one nanosecond, namely about 30 cm in the spatial domain, which is not enough to capture the fine-grained spatial variation. To solve this problem, the CIR accumulation technique was proposed in \cite{Moschevikin2016}. Specifically, DWM1000 modules were triggered by their local RF clocks in the experiment, and the CIRs between any two UWB nodes were sampled at slightly different times. We can accumulate the CIRs within a short duration and align the CIRs around the reported first peaks. The short duration is set as a sliding window of 50 consecutive CIR measurements. The stride of the sliding window is set as one in this paper. In this way, we can obtain an uneven and oversampled CIR, as shown by the gray dots in Fig. \ref{fig:BgDyn_CIR}, while the blue dots show the amplitude of CIR of a single measurement. The CIR measurements without/with a moving person (background/dynamic) are shown in Fig. \ref{fig:BgDyn_CIR}. The delay ($x$ axis) denotes the ToF difference between the ToF reflected by the moving person and the ToF of the transceiver.

\subsection{ToF Estimation}
\subsubsection{Challenges on Background Mitigation}
\begin{figure}[t]
    \centering
    \begin{minipage}{0.235\textwidth}
        \centering
        \setlength{\abovecaptionskip}{-0.1cm}
		\setlength{\belowcaptionskip}{-0.1cm}
        \includegraphics[width=1\textwidth]{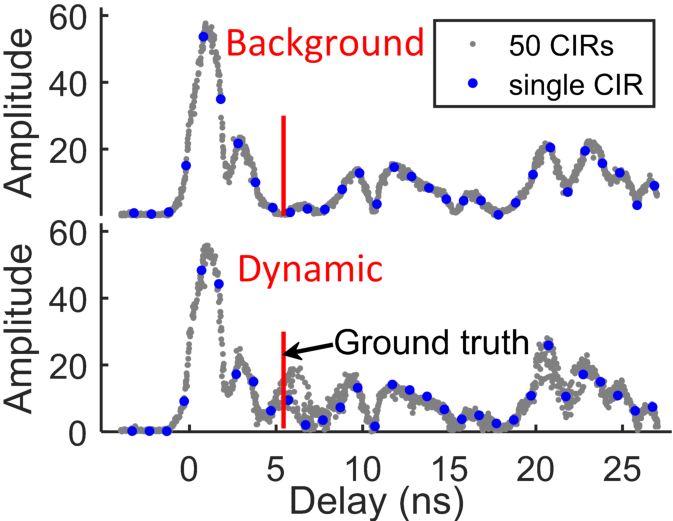} 
        \caption{Accumulated} CIRs for background and dynamic scenarios.
        \label{fig:BgDyn_CIR}
    \end{minipage}
    \hspace{0.1cm}
    \begin{minipage}{0.235\textwidth}
        \centering
        \setlength{\abovecaptionskip}{-0.1cm}
		\setlength{\belowcaptionskip}{-0.1cm}
        \includegraphics[width=1\textwidth]{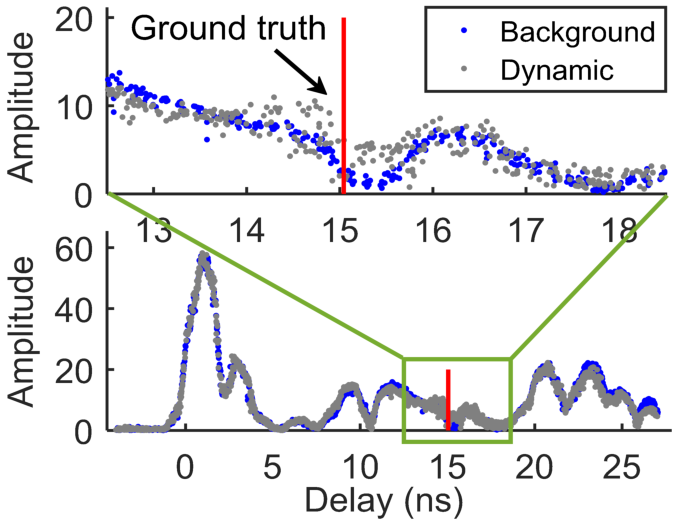} 
        \caption{Background and dynamic CIRs in case of a large delay.}
        \label{fig:BgDyn_CIR_far}
    \end{minipage}
\end{figure}

To obtain the reflected ToFs from the CIRs, it is necessary to mitigate the impact of the static line-of-sight (LoS) component and multipath components (MPCs) from the surroundings. An intuitive idea is the background subtraction \cite{Salmi2011}, however, this method is not effective in multipath scenarios\cite{McCracken2014}. Moreover, for the COTS DWM1000 module, the LoS bin of the CIRs is identified based on the LDE algorithm. So the accuracy of the LDE in the DWM1000 module also affects the background subtraction around the LoS component. To this end, the conventional background subtraction method is not suitable for the adopted COTS UWB devices. Another challenge is the weak signal strength reflected by the pedestrian in case of a large delay. According to the experimental results, if the reflected path from the moving person is much longer than the LoS link, the reflected signal strength is very weak and has no distinct increment compared with the background CIR, as shown in Fig. \ref{fig:BgDyn_CIR_far}. So it is difficult to distinguish the components of the human reflection merely based on the CIR's amplitude. Fortunately, we observe that the moving person causes distinct CIR fluctuations around the ground truth and the following CIR segments with larger delays, as shown in Figs. \ref{fig:BgDyn_CIR} and \ref{fig:BgDyn_CIR_far}. This phenomenon had also been observed in \cite{Ledergerber2020,Moschevikin2016}, which allows us to estimate the reflected ToF via detecting where the CIR fluctuations happen on the delay scale. Taking advantage of this observation, we propose two ToF estimation methods using the CIR-/variance-based CNN models. Instead of simple background subtraction as in \cite{Salmi2011}, we feed both the background and dynamic metrics into the model. In this way, the CNN can learn the differences between background and dynamic metrics and mitigate the impact of background internally.

\subsubsection{Input Features}
\begin{figure}[t]
\centering
\setlength{\abovecaptionskip}{-0.05cm}
\setlength{\belowcaptionskip}{-0.1cm}
\includegraphics[width=0.475\textwidth]{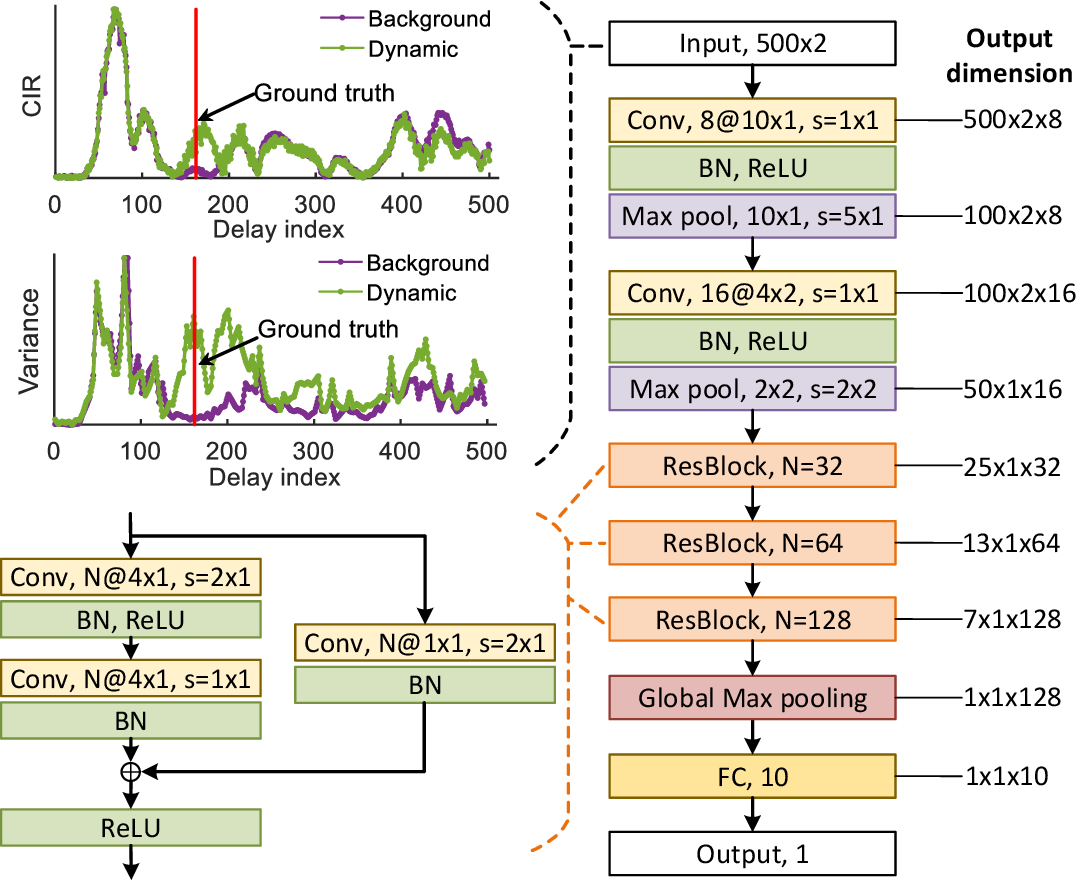} 
\caption{The designed CNN architecture for ToF estimation with the output dimension (right). The input of the CNN can be the normalized CIR or variance of background and dynamic scenarios (top left).}
\label{fig:CNNstruc}
\end{figure}

After performing the accumulation technique in Section III-A, we can obtain the oversampled CIR with a much higher resolution. As observed, the moving person will cause the CIR to fluctuate around the ground truth, which also means a larger CIR variance than the background case. To this end, we have established two CNN models for the reflected ToF estimation considering the different input features, namely the CIRs or the variances of the background and dynamic cases. Generally, to speed up the training progress and reduce the parameters of the neural networks, it is necessary to restrict the input size of CNN. On the other hand, the accumulated CIRs have different delay scales between each other. To this end, we resample the CIR linearly along with an even delay scale with, e.g., 500 samples (delay resolution 0.059 ns), and feed the resampled CIRs into the CNN model. Moreover, to ensure the effectiveness of variance, the number of samples within the sliding window along the CIR profile cannot be too small. To this end, we calculate the variance of CIR with a larger interval than the CIR resampling above. For example, the size of the variance series is set as 125 (delay resolution 0.236 ns). In this paper, we want to adopt the same CNN structure for either CIR or variance. So we interpolate the variance to the same size as CIR (i.e., 500) for later training and testing. An example of the resampled CIR and variance after normalization is presented in Fig. \ref{fig:CNNstruc}.

\begin{figure}[t]
\centering
\setlength{\abovecaptionskip}{-0.1cm}
\setlength{\belowcaptionskip}{-0.1cm}
\includegraphics[width=0.475\textwidth]{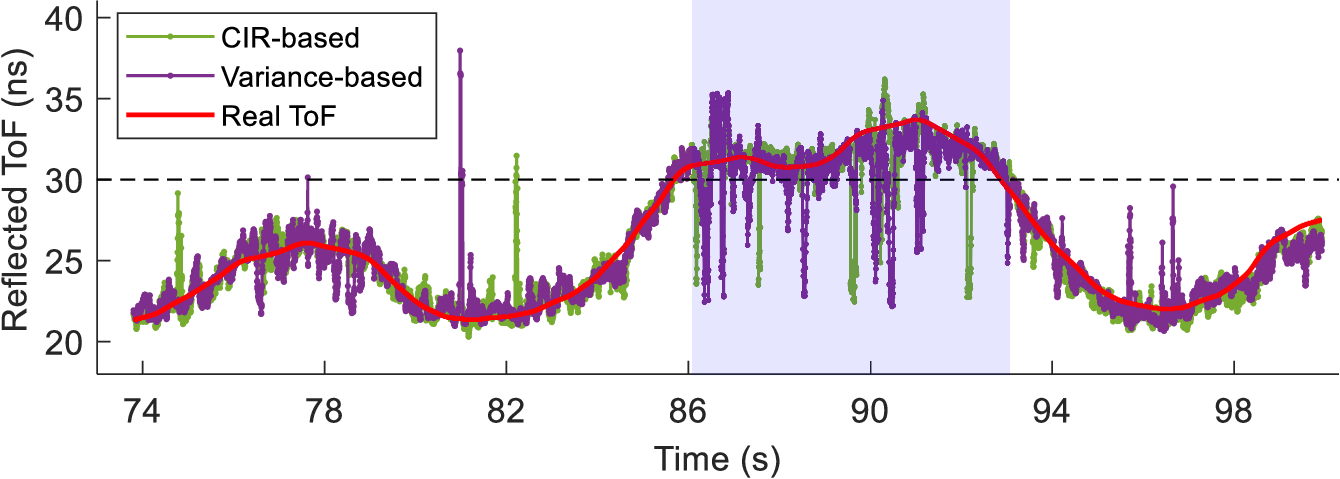} 
\caption{The estimated ToFs based on CIR- or variance-based CNN model.}
\label{fig:PredToF}
\end{figure}

\subsubsection{CNN Model}
When implementing the CNN, we model the reflected ToF estimation as a regression problem. The output of CNN is the delay index where the ground truth is located on the input CIR/variance profiles. The predicted delay index can easily be transformed to the reflected ToF by multiplying the delay resolution. In the designed CNN model, the elementary convolutional block is adopted: convolution (Conv), batch normalization (BN), activation (ReLU), and pooling layers. Especially, residual blocks (ResBlock) have been adopted to avoid the degradation problem of deep networks \cite{He2016}. The details of the proposed CNN's architecture and the corresponding output dimension of each layer are presented in Fig. \ref{fig:CNNstruc}. Adam optimizer \cite{kingma2017} has been adopted to train the weights of CNN. The batch size is set as 0.05 times of the training set. The maximum epoch is 100. Early stopping will be triggered to ensure generalization accuracy in case of no validation loss decrease for five epochs. The learning rate is set as 0.001. Fig. \ref{fig:PredToF} shows an example of the estimated ToFs based on the proposed CNN model versus the real ToFs, in which the mean absolute errors (MAEs) are less than 0.8 ns. Meanwhile, we can also observe that the proposed models have some outliers with distinct negative offsets when the ToFs are, e.g., larger than 30 ns, as the shadow rectangle shown in Fig. \ref{fig:PredToF}. This is because the signal strength is weak when the person moves far away from the transceiver link, the proposed models have some false alarms around the LoS components caused by LDE detection errors.
\subsection{Particle Filter-based Tracking}
\begin{figure}[t]
\centering
\setlength{\abovecaptionskip}{-0.15cm}
\setlength{\belowcaptionskip}{-0.05cm}
\includegraphics[width=0.49\textwidth]{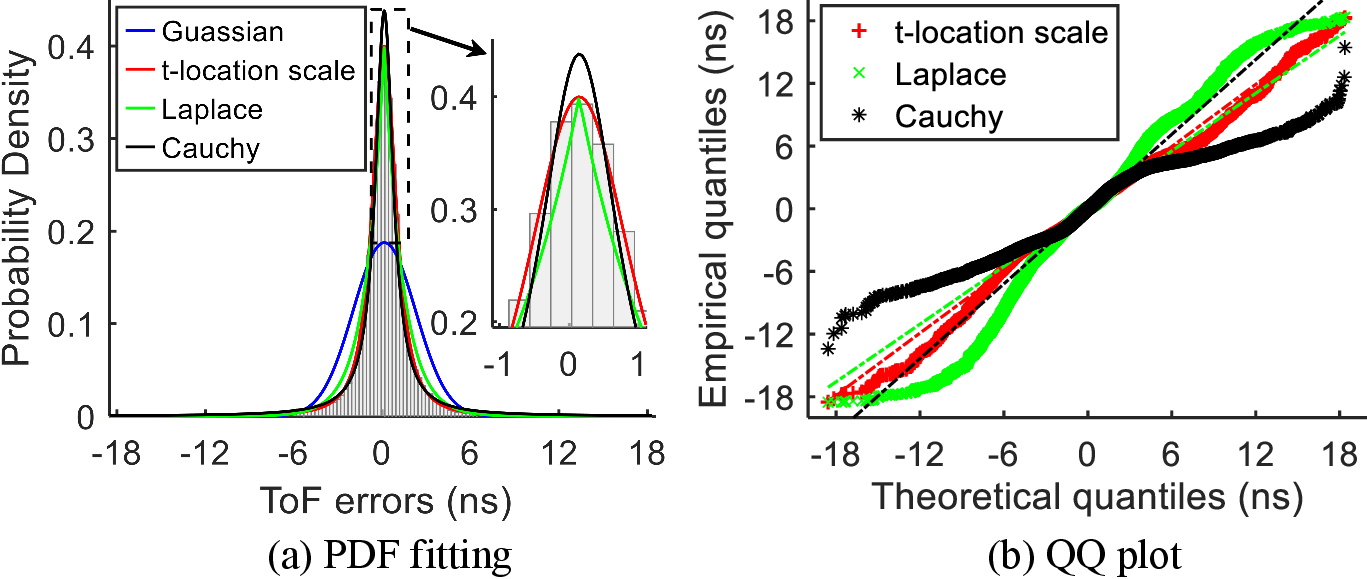}
\caption{PDF fitting to the reflected ToFs' errors and the QQ plots.}
\label{fig:PDF_fit}
\end{figure}

After obtaining the reflected ToFs of the multiple pairs of UWB links, we can derive the moving person's trajectory. Conventional elliptical localization is under the assumption of Gaussian ranging errors, which is sensitive to the outliers. In this paper, we adopt the particle filter (PF) algorithm for the tracking purpose \cite{Gustafsson2002}. Considering the computational cost of PF, we set the state as the 2-D coordinates $\mathbf{x}_{\rm{P}}$ of the moving person, which is updated via the simplified movement model $\mathbf{x}_{\rm{P}}^{(t+1)}=\mathbf{x}_{\rm{P}}^{(t+1)}+\Delta t\cdot{n}_v$, where $\Delta t$ is the time difference between $t$-th and $(t+1)$-th timestamps. ${n}_v$ is the Gaussian assumption of velocity. A set of $K$ particles (e.g., $K=200$) has been utilized to estimate the state to represent possible locations within the targeted area. We update the particles' weights via the probability density function (PDF) of the reflected ToFs' errors. To determine the PDF, we fit the histogram with four common distributions (Gaussian, Laplace, Cauchy, and $t$ location-scale distributions). As shown in Fig. \ref{fig:PDF_fit}(a), Laplace, Cauchy, and $t$ location-scale distributions have much better fitting performance than Gaussian distribution benefiting from the ability to model distributions with heavier tails. Fig. \ref{fig:PDF_fit}(b) presents the goodness-of-fit of Laplace, Cauchy, and $t$ location-scale distributions via quantile-quantile (QQ) plots. The plot (red plus signs) of $t$ location-scale distribution produces an approximately straight line compared to the other two plots, suggesting that the reflected ToFs' errors follow $t$ location-scale distribution in our case.

\begin{figure}[t]
\centering
\setlength{\abovecaptionskip}{-0.15cm}
\setlength{\belowcaptionskip}{-0.05cm}
\includegraphics[width=0.495\textwidth]{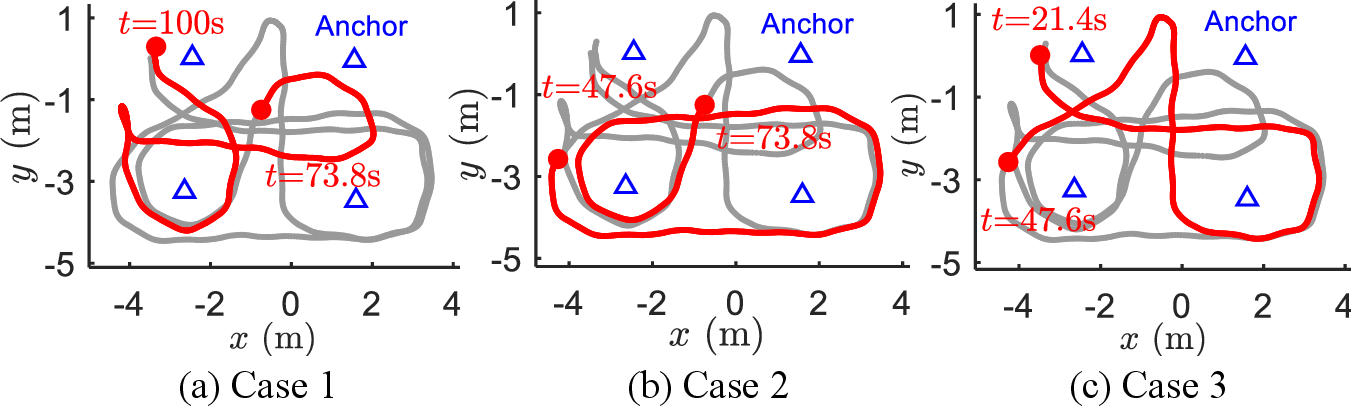}
\caption{Training (grey) and test (red) sets partitioning on the whole trajectory.}
\label{fig:TrainTestSet}
\end{figure}

\section{Performance Evaluation}
To validate the tracking accuracy of the proposed CNN-based algorithms, we divide the dataset into two parts, namely training and test sets. As mentioned in Section II, the dynamic period with a person moving is from 21.4 s to 100 s, namely 78.6 s duration. We have separated the training set as 52.4 s (about 66.7\% of the whole trajectory) and the test set as 26.2 s. Fig. \ref{fig:TrainTestSet} shows three cases of dataset partitioning. Four UWB nodes (blue triangles in Fig. \ref{fig:TrainTestSet}) are adopted in the datasets, so there are six pairs of TRx links. For the training set, we combine the CIR/variance series of all pairs of links. The codes for the ToF estimation and the passive tracking are open access to encourage further investigation\footnote{[Available]:\url{https://github.com/CLongLi/UWB-Radar-Pedestrian-Tracking}}.

\begin{figure}[t]
\centering
\setlength{\abovecaptionskip}{-0.15cm}
\includegraphics[width=0.495\textwidth]{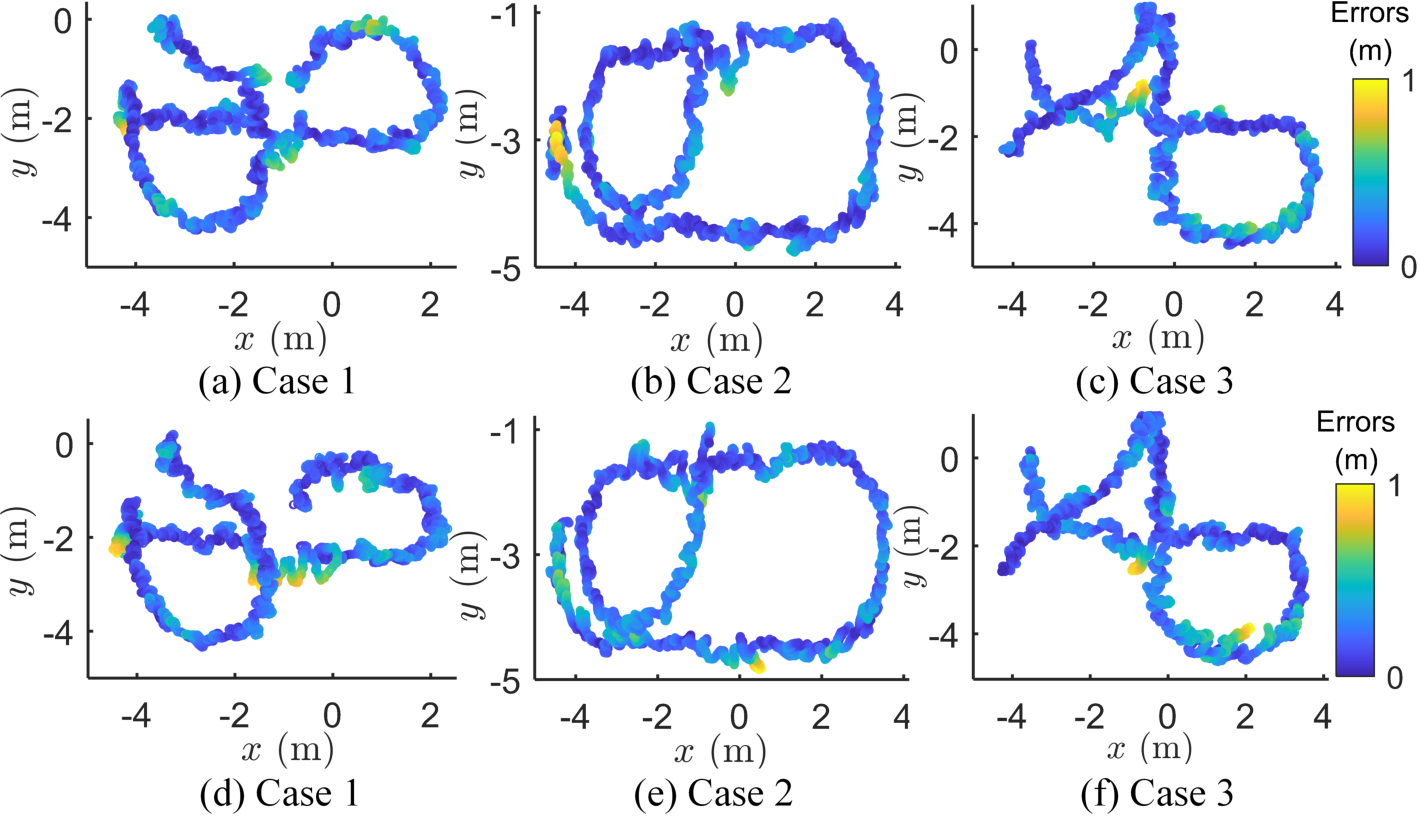}
\caption{Passive tracking results: (a)-(c) CIR-based. (d)-(f) Variance-based.}
\label{fig:TrackingResults}
\end{figure}

\begin{table}[t]
\centering
\vspace{-1em}
\caption{Comparison of UWB-based passive human tracking.}
\label{table:trackingRMSE}
\begin{tabular}{l||c|c|c}
\hline
\textbf{Methods}       & \textbf{RMSEs (cm)}     & \textbf{Improvement}& \textbf{Run time (ms)}\\
\hline\hline
\textbf{Method in}\cite{Ledergerber2020}&34.75-40.85&        /         &    1.71          \\ \hline
\textbf{CIR-based}     &     26.12-29.24            &    24.8-28.4$\%$ &    5.36          \\ \hline
\textbf{Variance-based}&     26.36-27.95            &    24.1-31.6$\%$ &    5.97          \\ \hline
\end{tabular}
\end{table}

\subsubsection{Tracking Accuracy}Fig. \ref{fig:TrackingResults} shows the passive tracking results of the three cases of datasets partitioning based on the proposed CNN methods. As presented in Table I, the CIR- and variance-based have achieved comparable accuracy with the root-mean-square errors (RMSEs) 26.12-29.24 cm and 26.36-27.95 cm, respectively. The CIR- and variance-based CNN models are also compared with the state-of-the-art method proposed in \cite{Ledergerber2020} which can be regarded as a variance-based LDE algorithm. As presented in Table \ref{table:trackingRMSE}, both of the proposed CIR- and variance-based CNN models outperform the method in \cite{Ledergerber2020} with more than 24.1\% improvement. Table \ref{table:trackingRMSE} also summarizes the average run time of each location estimation (including online ToF estimation and tracking). The running environment is Dell OptiPlex 7050 equipping Intel(R) Core(TM) i7-7700 CPU@3.60GHz and 16 GB RAM. The proposed methods have clearly larger computation time (5.36-5.97 ms) than the method in \cite{Ledergerber2020}. However, it is already very fast and sufficient for real-time pedestrian tracking.
\par
\begin{figure}[t]
\centering
\setlength{\abovecaptionskip}{-0.15cm}
\setlength{\belowcaptionskip}{-0.1cm}
\includegraphics[width=0.49\textwidth]{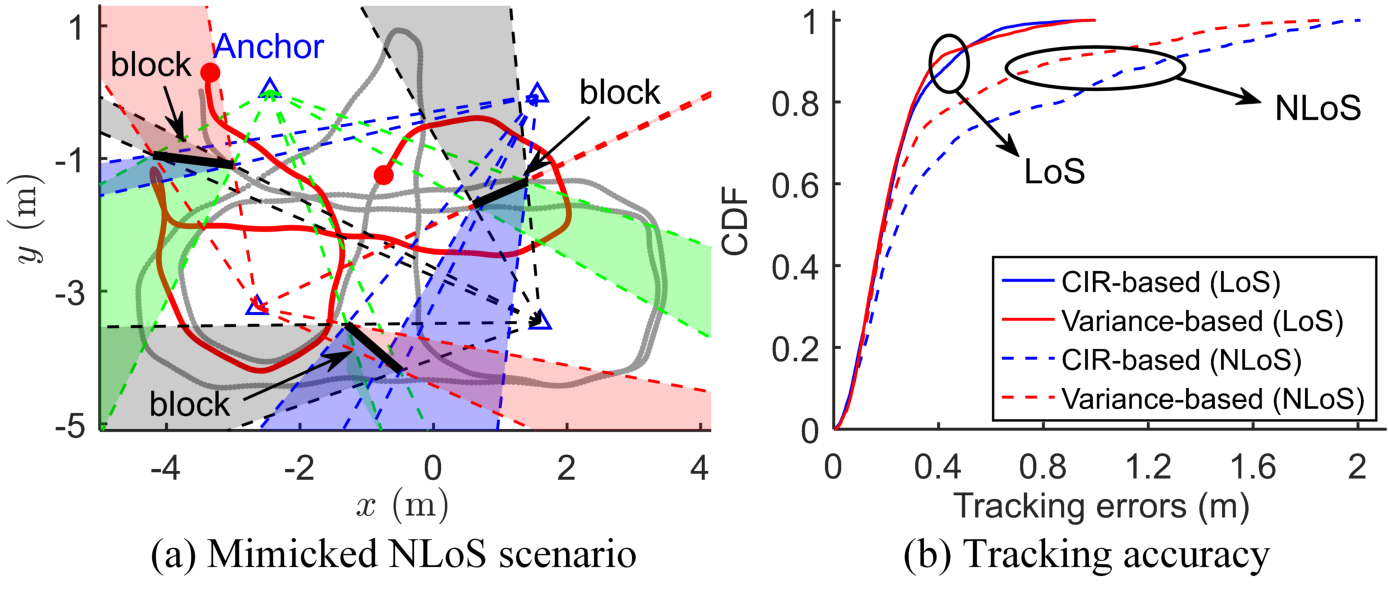}
\caption{Tracking accuracy evaluation for the NLoS scenario under \textit{Case 1}.}
\label{fig:NLoS_results}
\end{figure}

Non-line-of-sight (NLoS) brings challenges for the passive tracking, which generally should be avoided when deploying the UWB anchors. However, there may be cases that the direct link(s) between the moving person and the anchor(s) are blocked temporarily by, e.g., the furniture, which causes the signal reflected by the pedestrian undetectable. In Fig. \ref{fig:NLoS_results}(a), we mimic an NLoS scenario based on \textit{Case 1} data partitioning, where the pedestrian behind the blocks (shadow regions) cannot be detected by the corresponding transceiver. If the person moves to the shadow regions that are undetectable for more than one anchor, the pedestrian cannot be localized as only four anchors were adopted in the experiment. Fig. \ref{fig:NLoS_results}(b) compares the accuracy in case of LoS and NLoS using the cumulative distribution function (CDF) of the tracking errors. As expected, the NLoS degrades the tracking accuracy greatly despite the partial blocking as shown in Fig. \ref{fig:NLoS_results}(a). Furthermore, the variance-based CNN (with 51.41-cm RMSEs) outperforms the CIR-based CNN (with 66.43-cm RMSEs) in case of NLoS.

\subsubsection{Model Generalization}
\begin{figure}[t]
\centering
\setlength{\abovecaptionskip}{-0.15cm}
\setlength{\belowcaptionskip}{-0.1cm}
\includegraphics[width=0.495\textwidth]{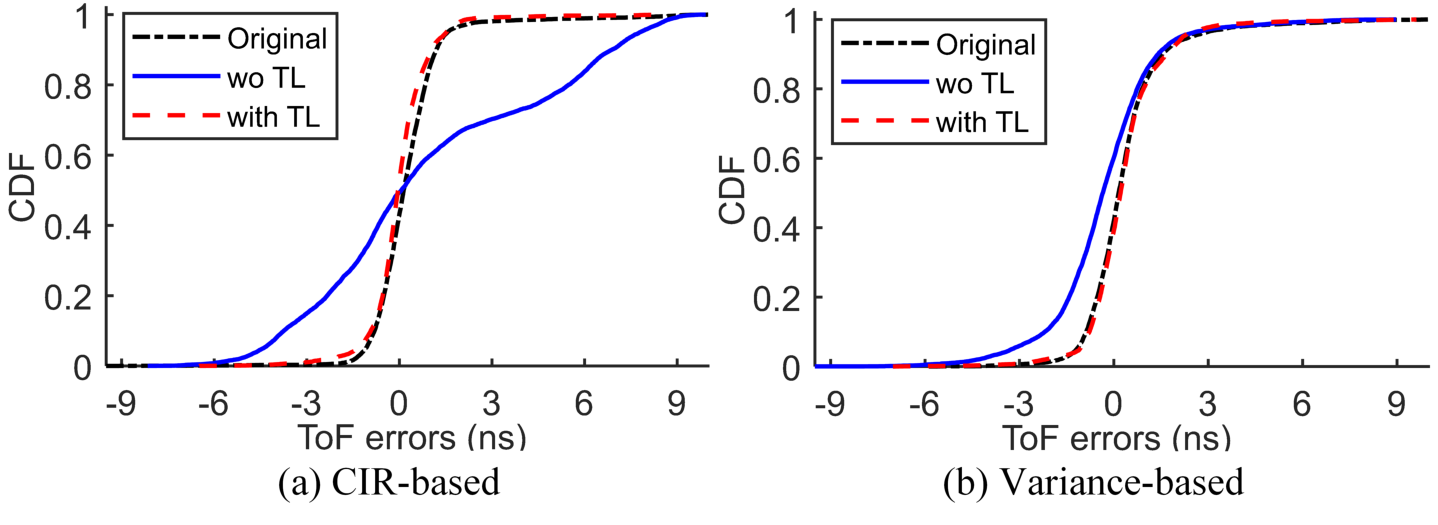}
\caption{ToF estimation accuracy with and without (wo) transfer learning.}
\label{fig:TransferToF}
\end{figure}

One of the major drawbacks of the deep learning-based algorithm is the weak generalization ability for different scenarios. Generally, the CIR of all TRx links is different from each other due to going through different propagation channels and having different interactions (reflections or scattering) with the surroundings. To a great extent, this is similar to the cases of the environment changes, which enables us to use the different CIR profiles to mimic the environment changing and investigate the impact of scenario changing on the proposed CNN models. We utilize the training set without the data from one of the TRx links (labeled as $\mathcal{A}$), to train the CNN models. The trained models are used to predict the reflected ToF of TRx link $\mathcal{A}$ in the test set. Besides, transfer learning (TL) has been adopted as the benchmark, which used 10\% training set of the TRx link $\mathcal{A}$ for TL's retraining. Fig. \ref{fig:TransferToF} shows the CDF plots of ToF errors based on CIR and variance models. The estimation accuracy of the CIR-based CNN model without TL (3.21-ns MAEs) is much lower than training using the whole training set of all pairs of links (labeled as \textit{Original} with 0.78-ns MAEs). In contrast, the variance-based CNN (with 1.25-ns MAEs) is slightly affected by the scenario changing as shown in Fig. \ref{fig:TransferToF}(b). It is clear that the variance-based CNN is more robust to campaign variation compared to the CIR-based CNN model, which is promising without the need of training from scratch or TL. 
\section{Conclusion}
In this paper, we have proposed a CNN-based passive human tracking method using the COTS UWB devices. The residual CNN models have been designed for both CIR- and variance-based ToF estimation. According to the experimental results, the proposed CIR- and variance-based CNN methods have achieved less than 30cm-RMSEs tracking accuracy. Especially, the variance-based CNN is more robust to NLoS scenario and is slightly environment-dependent than the CIR-based CNN model. The future works will consist of experimental evaluation of scenario variation, the maneuverable vehicles and pedestrian distinguishing, and multi-person tracking in the cluttered scenarios.
\section*{Acknowledgment}
This work is supported in part by the Excellence of Science (EOS) project MUlti-SErvice WIreless NETworks (MUSE-WINET), by the Research Foundation Flanders (FWO) SB Ph.D. fellowship under Grant 1SB7619N, and by the imec co-financed project UWB-IR AAA.
\balance
\bibliographystyle{IEEEtran}
\bibliography{UWBradar}

\end{document}